\documentclass[
 reprint,
nofootinbib,
prl,
 amsmath,amssymb,
 aps,
]{revtex4-2}

\usepackage{graphicx}
\usepackage{dcolumn}
\usepackage{bm}
\usepackage{hyperref}
\usepackage{xcolor}

\begin{document}


	\title{Nested active pointing control for interspacecraft laser interferometry}
	
	\author{Daikang Wei}
	\email{daikang.wei@aei.mpg.de}
	\author{Christoph Bode}%
    \author{Yihao Yan}
	\affiliation{%
		Max-Planck-Institut f\"ur Gravitationsphysik (Albert-Einstein-Institut), Callinstraße 38, 30167 Hannover, Germany
	}%
	\affiliation{%
		Leibniz Universit\"at Hannover, Institut für Gravitationsphysik, Callinstraße 38, Hannover 30167, Germany
	}%
	\author{Vitali M\"{u}ller}%
	\affiliation{%
		Max-Planck-Institut f\"ur Gravitationsphysik (Albert-Einstein-Institut), Callinstraße 38, 30167 Hannover, Germany
	}%
	\affiliation{%
		Leibniz Universit\"at Hannover, Institut für Gravitationsphysik, Callinstraße 38, Hannover 30167, Germany
	}%
	\author{Juan Jos\'e Esteban Delgado}%
	\author{Gerhard Heinzel}%
	\affiliation{%
		Max-Planck-Institut f\"ur Gravitationsphysik (Albert-Einstein-Institut), Callinstraße 38, 30167 Hannover, Germany
	}%
	\affiliation{%
		Leibniz Universit\"at Hannover, Institut für Gravitationsphysik, Callinstraße 38, Hannover 30167, Germany
	}%
	
	\date{\today}
\begin{abstract}
Precise pointing control is critical for interspacecraft laser interferometry, as angular misalignment introduces measurement noise and can even lead to laser link loss. We present a nested control architecture that uses differential wavefront sensing signals to drive a fast steering mirror (FSM) to track the incoming beam, while feeding the FSM deflections back to the attitude and orbit control system to suppress angle-dependent optical path variations. This scheme is experimentally validated in our hexapod-based setup. Relative to standalone FSM actuation, the nested configuration reduced the low-frequency root-mean-square pointing jitter by approximately a factor of 2 over 3--70 $\mathrm{mHz}$ in both directions. Additionally, tilt-to-length coupling was suppressed by an order of magnitude below 5 mHz and by two orders of magnitude below 0.5 mHz. These results demonstrate the feasibility of nested active pointing control for future interspacecraft laser interferometry missions.
\end{abstract}

\maketitle


\textit{Introduction.}---Spaceborne precision measurement has emerged as a critical method for studying the gravity field and fundamental physics. Interspacecraft laser interferometry was successfully demonstrated in the Gravity Recovery and Climate Experiment Follow-On (GRACE-FO) mission \cite{abich2019orbit,kornfeld2019grace}. The laser ranging interferometer (LRI) achieved nanometer-level ranging measurements between two satellites \cite{abich2019orbit, sheard2012intersatellite}, providing unprecedented insights into Earth's time-variable gravity field and climate dynamics \cite{tapley2004gravity, tapley2004grace, kornfeld2019grace}. Future GRACE-like missions, such as GRACE-Continuity (GRACE-C) and the Next-Generation Gravity Mission (NGGM), will also employ similar technology \cite{landerer2024towards,haagmans2020esa,nicklaus2022towards}. Beyond geodesy, the upcoming Laser Interferometer Space Antenna (LISA) mission aims to detect gravitational waves in the millihertz band by monitoring the relative distance variation between three spacecraft separated by millions of kilometers \cite{LISA}. Moreover, similar missions utilizing interspacecraft laser interferometry include the Chinese Taiji and TianQin projects \cite{Luo2016, Luo2020}, and the Japanese DECi-hertz Interferometer Gravitational wave Observatory (DECIGO) \cite{kawamura2011japanese}.

In interspacecraft interferometry, angular misalignment between the receiving (RX) and transmitting (TX) beams risks link loss due to reduced heterodyne efficiency \cite{Muller2017PhD,wanner2014analytical} and introduces tilt-to-length (TTL) coupling, a significant noise source in precision interferometry \cite{wegener2020tilt, hartig2022geometric, hartig2023non}. In the GRACE-FO LRI, differential wavefront sensing (DWS) signals were fed to a fast steering mirror (FSM) on the optical bench to coalign the local (LO) and TX beams with the RX beam, achieving a pointing stability below 10 µrad$\sqrt{\mathrm{Hz}}$ \cite{sheard2012intersatellite,schutze2014laser}. The spacecraft attitude is independently maintained by the attitude and orbit control system (AOCS) in fine-pointing mode, with pitch and yaw control better than $0.25~\mathrm{mrad}$ using star-tracker and inertial-measurement-unit data \cite{cossavella2022attitude}. For the LISA mission, which requires pointing noise below 10 nrad/$\sqrt{\mathrm{Hz}}$, DWS angular readouts will be directly used to control the spacecraft's attitude and telescope pointing \cite{LISA,hasselmann2025validation, gath2007drag}. In addition, a point-ahead angle mechanism will compensate for the offset angle between the RX and TX beams for each spacecraft \cite{langer2009design,houba2022lisa}.

Although the FSM effectively suppresses residual beam pointing jitter with respect to the line of sight (LOS) in the GRACE-FO LRI, it does not correct the spacecraft attitude itself and therefore cannot eliminate angle-dependent optical-path variations. These variations modify the detected interference phase and DWS signals, inducing TTL coupling and affecting pointing stability. Rotating the entire optical bench to track the incoming RX beam, like the LISA mission, would eliminate this non-stationary coupling but requires a drag-free and attitude control system  driven by micropropulsion \cite{LISA, gath2007drag, vidano2024drag}. Compared to the relatively quiet heliocentric orbit of LISA, spacecraft in low geocentric orbits experience more disturbed environments, where relying solely on a DWS-based AOCS for pointing stabilization can result in residual high-frequency jitter that compromises interspacecraft laser interferometric links.

Here, we propose a new precision-pointing control strategy for interspacecraft laser interferometry. The DWS signal, which carries coalignment information between the RX and TX beams \cite{heinzel2020tracking,wanner2012methods}, drives an FSM on the optical bench to compensate for pointing jitter with a unity-gain bandwidth of approximately 100 Hz. In parallel, the spacecraft attitude angle (SAA), inferred from the FSM angular deflections, is fed back to the AOCS to suppress non-stationary coupling caused by SAA dependence. The SAA, equivalent to the RX-beam incidence angle at the optical-bench aperture, is defined in the \textit{Angle definition} section of the End Matter. Further details regarding the new pointing control scheme are provided in the \textit{Concept} section of the End Matter.

This Letter reports an experimental study of nested active pointing control for interspacecraft laser interferometry. Under hexapod-driven jitter, the scheme achieves $\sim100~\mathrm{nrad}/\sqrt{\mathrm{Hz}}\times\mathrm{NSF}(f)$ pointing noise in both axes and markedly reduces TTL noise relative to DWS--FSM control alone, where $\mathrm{NSF}(f)$ is the noise-shaping function defined in the End Matter.

\begin{figure*}
	\centering
	\includegraphics[width=0.82\textwidth]{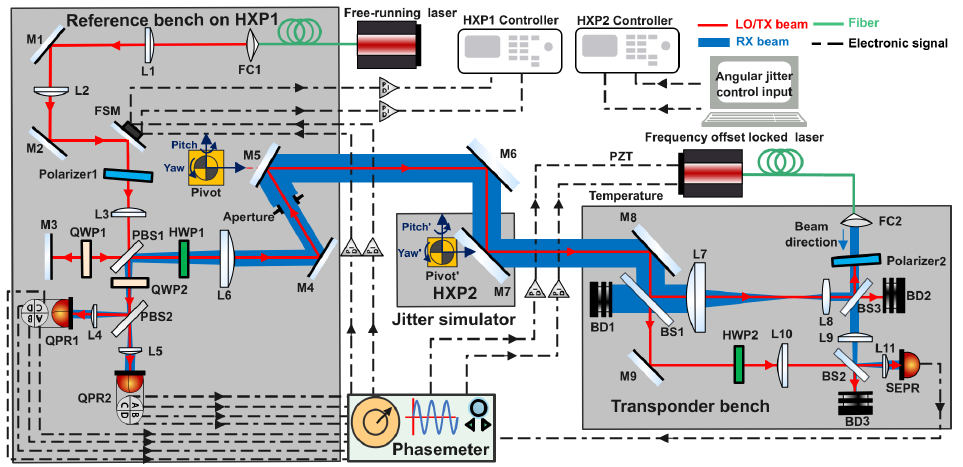}
	\caption{Schematic of the experimental setup. The hexapods (HXP1 and HXP2) used in this experiment are identical models (Newport, HXP100-MECA). L, lens; M, mirror; FC, fiber collimator; FSM, fast steering mirror;  HXP, hexapod; HWP, half-wave plate; SEPR, single-element photoreceiver; QWP, quarter-wave plate; QPR, quadrant photoreceiver; BS, beamsplitter; PBS, polarizing beamsplitter; BD, beam dump; RP, reference point; PZT, piezoelectric transducer.}
	\label{Exp_setup}
\end{figure*}

\textit{Methods.}---We consider a simplified one-way ranging scenario in which the RX beam enters the spacecraft optical bench, which exhibits attitude yaw angle $\alpha$ and pitch angle $\beta$. The misalignment angles between the TX and RX beams are denoted as $\alpha^{\prime}$ and $\beta^{\prime}$. The ranging measurement comprises contributions from the optical bench path length $\rho_{OB}\left( \alpha,\beta, \alpha^{\prime}, \beta^{\prime} \right)$, and the interspacecraft path length $\rho_{S/C}$, as expressed as
\begin{equation}
    \rho\left( \alpha,\beta, \alpha^{\prime}, \beta^{\prime} \right) = \rho_{OB}\left( \alpha,\beta, \alpha^{\prime}, \beta^{\prime} \right) + \rho_{S/C}.
    \label{phase_ranging}
\end{equation}
Without active pointing loops, the SAA ($\alpha$, $\beta$) equals the misalignment angle ($\alpha^{\prime}$, $\beta^{\prime}$). To achieve antiparallel alignment between the RX and TX beams on the optical bench, the FSM deflects the TX and LO beams in response to the feedback from DWS signals, which are defined here as DWS-FSM loops. These loops suppress $\alpha^{\prime}$ and $\beta^{\prime}$ to zero, thereby mitigating the TTL coupling arising from angles $\alpha^{\prime}$ and $ \beta^{\prime}$. Within the DWS-FSM loop bandwidth, the FSM angular deflections $\gamma$ and $\eta$ are linearly related to the SAA $\alpha$ and $\beta$, respectively. We introduce additional loops, called SAA-AOCS loops, based on $\gamma$ and $\eta$ that drive the AOCS to maintain $\alpha$ and $\beta$ near zero. The combination of DWS-FSM and SAA-AOCS loops further mitigates the TTL coupling from angles $\alpha$ and $\beta$. The DWS offset, the residual signal at perfect beam alignment, varies with the SAA. This variation is suppressed by SAA-AOCS loops, enabling the DWS-FSM loops to operate at a constant offset and further enhance pointing stability.

The experimental setup is shown in Fig.~\ref{Exp_setup}. The interferometric link was established between the reference bench and the transponder bench. The jitter simulator, a reflective mirror (M7) mounted on a hexapod (HXP2), was positioned between these benches. The simulator's principle is to control the hexapod's rotation to tilt the mirror, thereby inducing angular variations in the RX beam that replicate the attitude jitter of the reference spacecraft. The phase information and the DWS signal were measured by a phasemeter, which is based on an all-digital phase locked loop (ADPLL) technique to track the microradian phase variation \cite{shaddock2006overview,gerberding2013phasemeter}.

As a modification of the on-axis LRI optical bench \cite{yang2022axis, wei_2026}, the reference bench was mounted on a hexapod (HXP1) capable of six-degrees-of-freedom motion to emulate the AOCS. A 5 mW free-running laser at 1064 nm was delivered via a fiber, lenses (L1 and L2), and mirrors (M1 and M2), with its beam waist located on an FSM (Physik Instrumente, S-325). After the FSM, a linear polarizer was used to generate a polarized beam which can be decomposed into a 5\% parallel-polarized (P-pol) component and a 95\% perpendicular-polarized (S-pol) component. The P-pol beam passed through a polarizing beam splitter (PBS1) and interfered with the RX beam as the LO beam. The S-pol beam was converted to P-pol after passing through a quarter-wave plate (QWP1) twice and sent to the transponder bench as the TX beam. The FSM was imaged onto quadrant photoreceivers QPR1 and QPR2 using two-lens systems (L3 and L4; L3 and L5), respectively. The aperture between mirrors M4 and M5 was imaged onto QPR1 and QPR2 as the conjugate plane of the FSM. The pivot point of HXP1 was aligned with the mirror image of the aperture center.

On the transponder bench, a portion of the laser was directed to a single-element photoreceiver (SEPR) to serve as a reference beam, producing a heterodyne interference signal with the TX beam from the reference bench. The heterodyne frequency, measured by a phasemeter, was fed back to control the laser's temperature and cavity length, achieving a 7.3 MHz frequency-offset lock to the incoming beam. Another portion of the laser was expanded to a waist radius of approximately 21 mm and sent to the reference bench as the RX beam. The beam was clipped by the aperture to a power of about 100 nW, approximating a flat-top profile characteristic of the far-field beam from a distant spacecraft.

\begin{figure}
	\centering
	\includegraphics[width=0.483\textwidth]{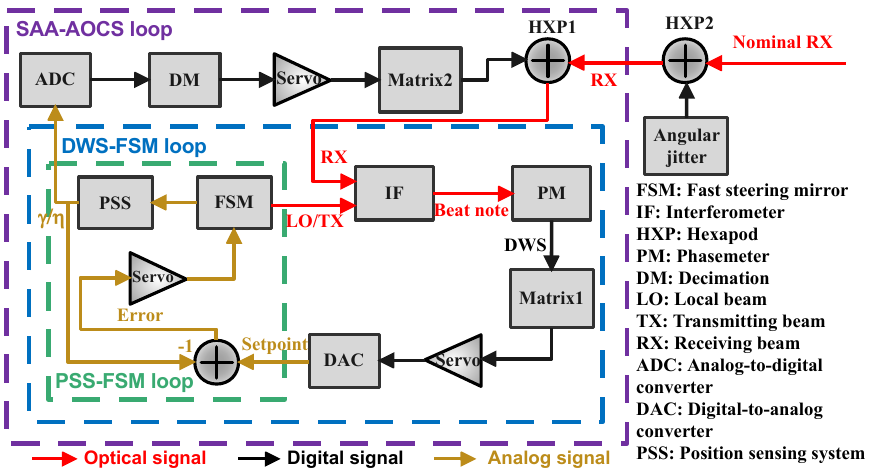}
	\caption{Block diagram of the nested pointing control architecture. Matrix1 and Matrix2 transform the rotational coordinate frames from the QPR to the FSM and from the FSM to the hexapod, respectively. The yaw and pitch loops are identical, so only one axis is depicted.}
	\label{Pointing_loops}
\end{figure}

The implemented control scheme is illustrated in Fig.~\ref{Pointing_loops}. In addition to the two DWS-FSM control loops in yaw and pitch directions, this scheme incorporates two additional SAA-AOCS loops. Attitude jitter tilts the RX beam, producing nonzero DWS signals ($\mathrm{DWS_h}$ and $\mathrm{DWS_v}$) that drive the FSM to maintain coalignment between the RX and TX beams. Meanwhile, the FSM deflection angles $\gamma$ and $\eta$, measured by the position sensing system (PSS), are fed back to the AOCS---implemented by a hexapod here---to suppress SAA variations. To ensure accurate tracking of the commanded FSM deflections, the FSM operates in closed loops with the PSS: the measured deflections are compared with their commanded values, and the resulting errors drive the FSM controller, forming the innermost and fastest PSS--FSM loops, as indicated by the green dashed box in Fig.~\ref{Pointing_loops}. The PSS comprises strain-gauge sensors integrated into the FSM with a deflection-angle resolution of $100~\mathrm{nrad}$. Overall, the FSM rapidly corrects beam misalignment via DWS feedback, whereas the SAA-AOCS loops operate on a slower time scale to suppress SAA variations and keep the FSM operating near its central position.

In the experiment, the DWS signals from QPR1 provided feedback to the FSM through the DWS-FSM loops, while the PSS-measured angles were decimated to 2.38 Hz for hexapod actuation through the SAA--AOCS loops. Frame discrepancies among QPR1, the FSM, and the hexapod, caused by misalignments, were corrected by two calibrated transformation matrices (Matrix1 and Matrix2). Open-loop transfer function measurements showed unity-gain bandwidths of 112.77 Hz and 98.04 Hz for the horizontal and vertical DWS-FSM loops, respectively, while the SAA-AOCS loops achieved bandwidths of 67 mHz and 62 mHz. More details about the transfer functions are provided in the End Matter. The DWS-FSM loops maintained lock throughout the motions of the reference bench and the jitter simulator.

\begin{figure}
	\centering
	\includegraphics[width=0.48\textwidth]{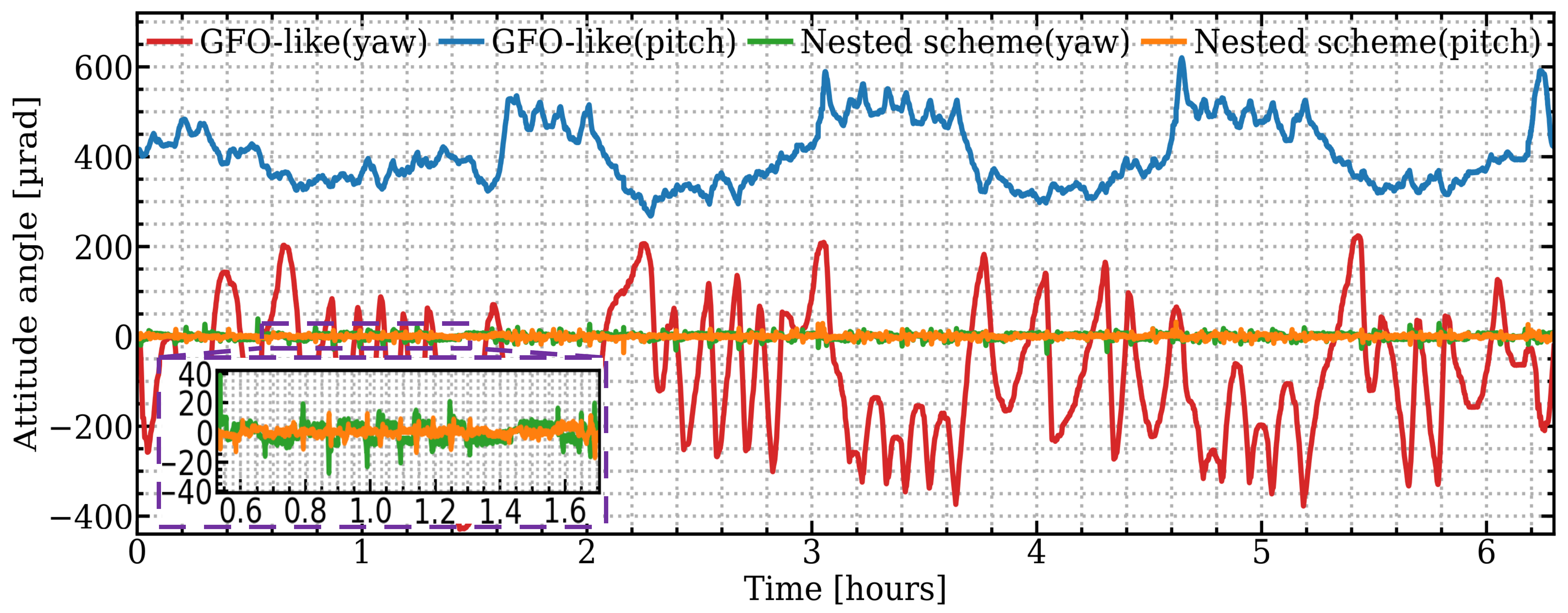}
	\caption{Measured attitude angles as a function of time, inferred from the FSM-PSS measurements using Eq.~\ref{fsm_to_attitude}. Yaw and pitch correspond to the horizontal and vertical SAAs, respectively. The curves labeled GFO-like were obtained in the GRACE-FO-like DWS-FSM configuration, whereas those labeled nested scheme were obtained in separate measurements with both the DWS-FSM and SAA-AOCS loops closed.}
	\label{SAA_angle}
\end{figure}

\textit{Results.}---In this experiment, attitude jitter was simulated using GRACE-FO pointing data derived from star-camera measurements acquired on January 1, 2019 \cite{goswami2021analysis}. With the attitude jitter, two independent measurements were performed: one in the GRACE-FO-like control, in which the residual  jitter is compensated only by the DWS-FSM loops, and the other using the proposed nested scheme. The corresponding SAAs, inferred from the FSM-PSS measurements using Eq.~\ref{fsm_to_attitude} defined in End Matter, are shown in Fig.~\ref{SAA_angle}. With the GRACE-FO-like control, the FSM quickly compensated for the beam misalignment induced by the attitude jitter. As a result, the SAAs followed the injected attitude jitter, as shown by the red and blue curves. With the SAA-AOCS loops additionally closed, the attitude variations were suppressed to within approximately $\pm$40 µrad, as indicated by the green and orange curves, with the residual level limited by the motion precision of the hexapod. Consequently, in contrast to the large FSM excursions in the GRACE-FO-like configuration, the FSM remained close to its central position, with only small deflections about zero.

\begin{figure}
	\centering
	\includegraphics[width=0.483\textwidth]{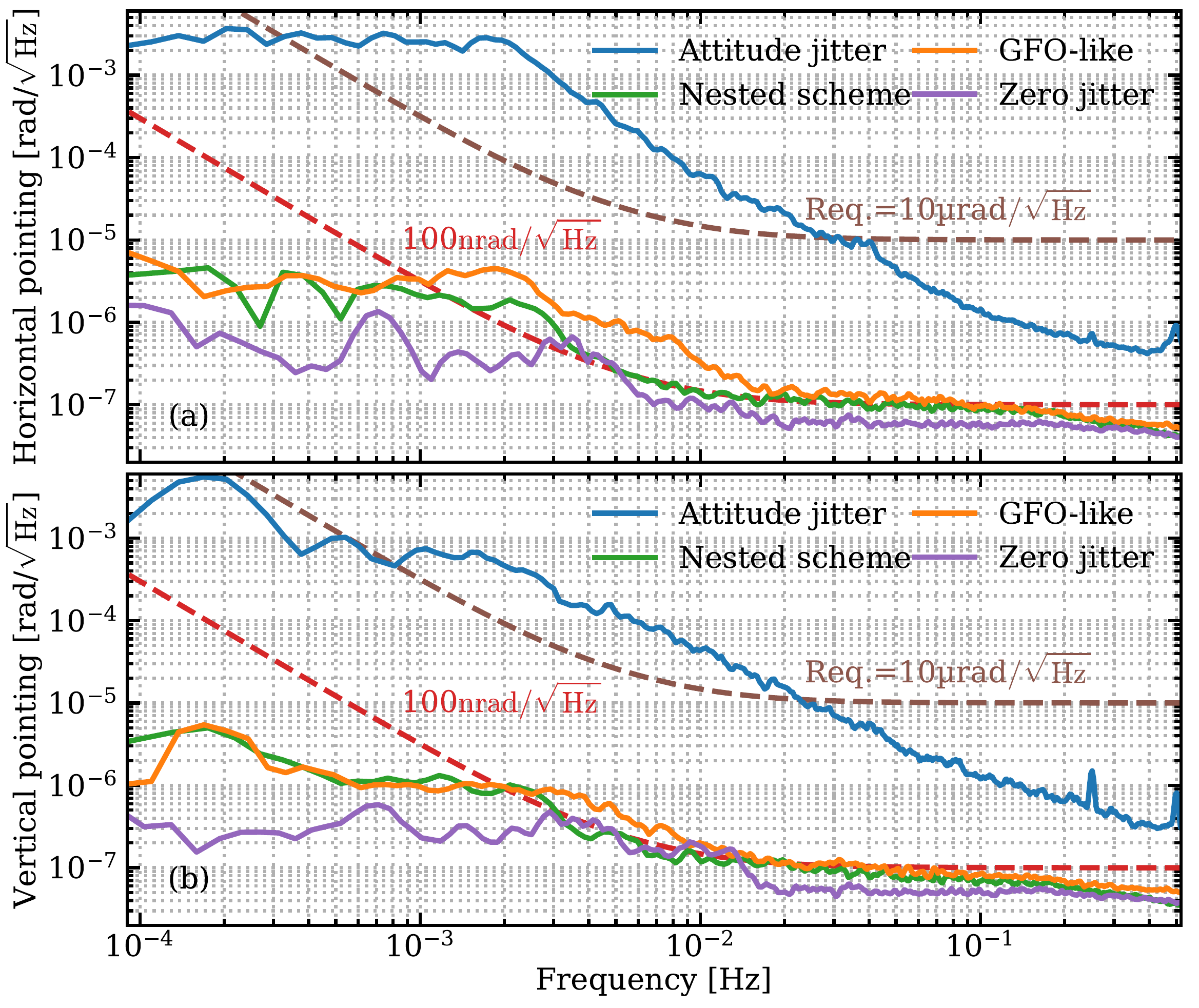}
	\caption{Amplitude spectral densities of the (a) horizontal and (b) vertical pointing angles. The blue curves show the attitude jitter generated by HXP2 rotations and inferred from the FSM-PSS measurements using Eq.~\ref{fsm_to_attitude}, which represents the pointing performance with the DWS-FSM and SAA-AOCS loops open. The GFO-like curves correspond to DWS-FSM control alone, as in the GRACE-FO LRI, whereas the nested scheme curves include both DWS-FSM and SAA-AOCS control. For the zero-jitter measurement, HXP2 remained static.}
	\label{Measured_pointing}
\end{figure}

Using an experimentally determined conversion matrix, out-of-loop DWS signals from QPR2 were converted into pointing angles. The corresponding amplitude spectral densities (ASDs) for the horizontal and vertical pointing angles are presented in Fig.~\ref{Measured_pointing}(a) and (b), respectively. The open-loop pointing jitter, denoted by blue curves, was suppressed by the DWS-FSM loops to the stability levels depicted by the orange lines. In both directions, pointing stability improved by two orders of magnitude at frequencies below 20 mHz. 

The SAA--AOCS loops further suppressed low-frequency pointing noise to $\sim100~\mathrm{nrad}/\sqrt{\mathrm{Hz}}\times\mathrm{NSF}(f)$ in both axes, approximately two orders of magnitude below the NGGM requirement of 10 µrad$/\sqrt{\mathrm{Hz}}\times\mathrm{NSF}(f)$ \cite{nicklaus2020laser}. Compared with DWS-FSM loops alone, the root-mean-square (RMS) residual jitter was reduced by a factor of 2-3 over $3$--$70~\mathrm{mHz}$ horizontally, with a slightly smaller reduction vertically. Below $3~\mathrm{mHz}$, both configurations were limited by thermal fluctuations. The remaining gap to the zero-jitter performance was limited by the finite precision of the hexapod \cite{NewportHXPManual} and the bandwidth of the SAA-AOCS loops.

\begin{figure}
	\centering
	\includegraphics[width=0.483\textwidth]{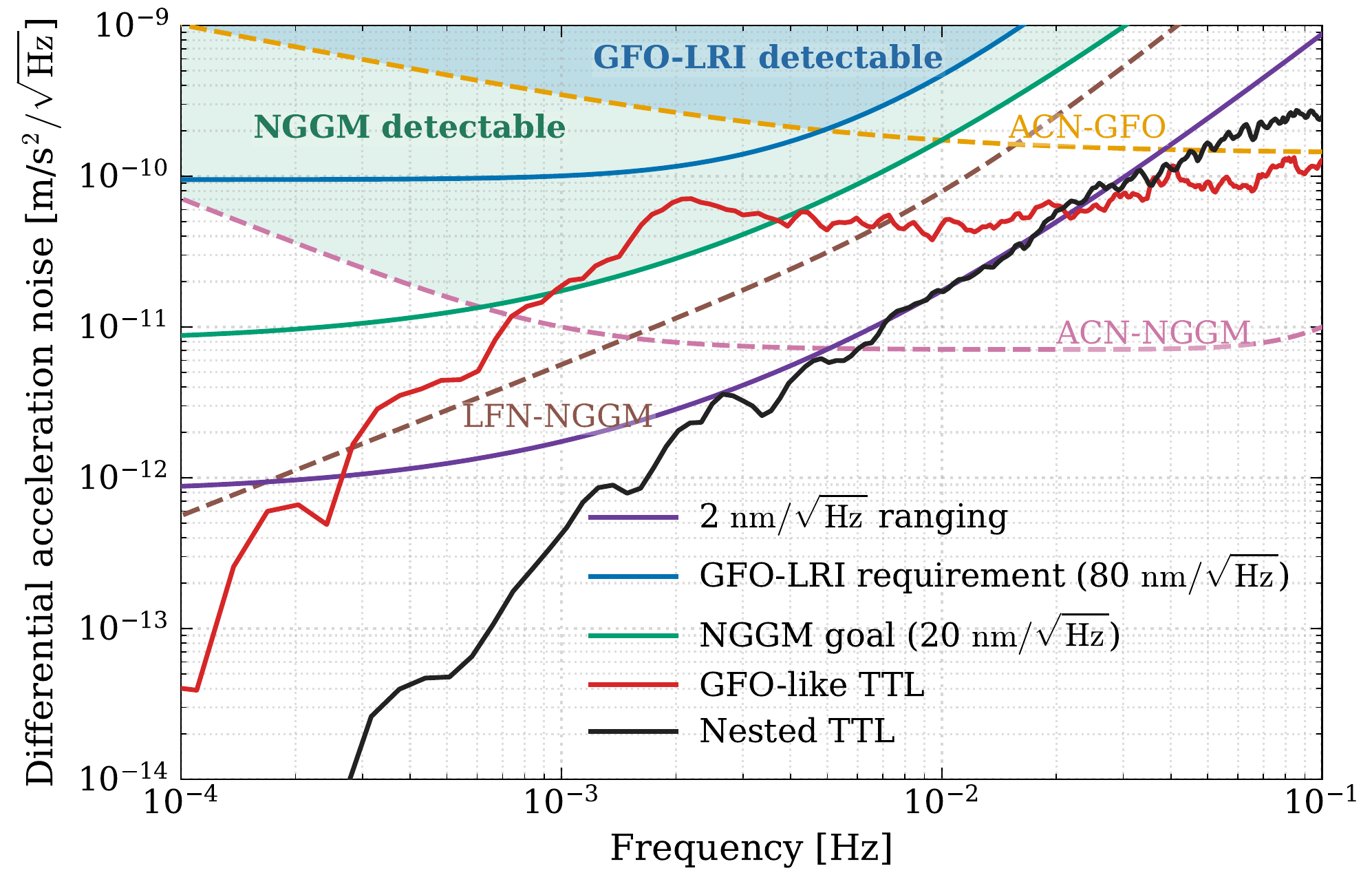}
	\caption{Amplitude spectral density of equivalent differential acceleration noise. The solid purple, blue, and green curves show the equivalent noise obtained from ranging-noise levels of $2~\mathrm{nm}/\sqrt{\mathrm{Hz}}\times\mathrm{NSF}(f)$, $80~\mathrm{nm}/\sqrt{\mathrm{Hz}}\times\mathrm{NSF}(f)$ (GRACE-FO LRI), and $20~\mathrm{nm}/\sqrt{\mathrm{Hz}}\times\mathrm{NSF}(f)$ (NGGM), respectively \cite{abich2019orbit,nicklaus2022towards}, computed from the second derivatives of the corresponding ranging quantities. The red and black curves show the TTL contributions estimated from the measured attitude angles using GRACE-FO LRI coupling factors for the GRACE-FO-like and nested configurations, respectively \cite{wegener2020tilt}. Dashed curves denote the projected NGGM laser-frequency noise (LFN-NGGM) \cite{nicklaus2020laser}, GRACE-FO accelerometer noise (ACN-GFO) \cite{darbeheshti2017instrument}, and NGGM accelerometer noise (ACN-NGGM) \cite{massotti2021next,QSG4EMTConsortium2024}. The blue and green hatched regions indicate signals above both the ranging and accelerometer noise floors for GRACE-FO LRI and NGGM, respectively.}
	\label{estimated_TTL_diff_acc}
\end{figure}

To first order, the TTL-induced ranging variation is 
\begin{equation}
\delta L(\alpha,\beta)\approx
a\alpha+b\beta,
\label{lps}
\end{equation}
where $a$ and $b$ are the yaw and pitch TTL coupling factors. Because the GRACE-FO-LRI TTL is dominated by the offset between the LRI reference point and the spacecraft center of mass \cite{wegener2020tilt}, this linear model is sufficient for the small attitude excursions considered here. We applied the GRACE-FO LRI coupling factors \cite{wegener2020tilt} to the measured attitude angles in Fig.~\ref{SAA_angle} to estimate the TTL-induced ranging noise for the GRACE-FO-like and nested configurations. The corresponding equivalent differential acceleration noise, a key metric of payload sensitivity in GRACE-like missions, is shown by the red and black curves, respectively, in Fig.~\ref{estimated_TTL_diff_acc}.

In the GRACE-FO-like configuration, SAA fluctuations couple into the ranging signal. The nested SAA-AOCS control suppresses these fluctuations at low frequencies, reducing the TTL contribution by at least one order of magnitude below 5 mHz and by at least two orders below 0.5 mHz. Because the injected attitude jitter is concentrated mainly below 40 mHz, the SAA-AOCS loops introduce excess attitude noise above this frequency, leading to a modest increase in TTL coupling above 40 mHz.

The unity-gain frequencies of the SAA-AOCS loops are limited by the hexapod's response speed and positional precision. Nevertheless, these bandwidths represent realistic orders of magnitude for the bandwidths of actual satellite AOCS pointing control. With improved hexapod angular motion precision and higher unity-gain frequencies, the degradation of TTL coupling at higher frequencies could be mitigated.

The impact on the geodesy observable is most evident over 0.7--7.5 mHz. In the GRACE-FO-like configuration, the TTL contribution exceeds the projected NGGM laser frequency noise (LFN) floor at the upper part of this band and the accelerometer noise (ACN) floor at the lower part. With the nested scheme, the TTL contribution is reduced below both limits, so that, with all other noise unchanged, the achievable differential-acceleration sensitivity is limited by the LFN and ACN. The corresponding residual noise remains below $2~\mathrm{nm}/\sqrt{\mathrm{Hz}}\times\mathrm{NSF}(f)$, about an order of magnitude below the NGGM ranging-noise goal \cite{nicklaus2022towards}. Consequently, further improvements in LFN and ACN could translate directly into improved gravity-field sensitivity, without TTL coupling becoming the limiting factor.

Beyond optical TTL coupling, improved spacecraft attitude stability may also benefit the differential-acceleration observable by reducing attitude-dependent gravity-gradient, rotational, and projection couplings in the accelerometer measurement \cite{cesare2022drag}.

\textit{Discussion and Conclusions.}---We have demonstrated nested active pointing control for interspacecraft laser interferometry using a transponder-based link spanning three optical benches. GRACE-FO-like attitude jitter was injected by a jitter simulator, while DWS signals drove the FSM to maintain RX-TX beam alignment. Simultaneously, the measured FSM deflections were fed back to the hexapod to rotate the reference bench, suppressing SAA variations and nonstationary coupling in the DWS signal and measured path length.

The DWS-FSM loops achieved unity-gain frequencies of 112.77 Hz (horizontal) and 98.04 Hz (vertical), well above the characteristic spacecraft attitude-jitter frequencies below 1 Hz \cite{daniel_intersatellite_2015}. The slower SAA-AOCS loops, with unity-gain frequencies of 67 and 62 mHz, suppressed low-frequency attitude variations. Out-of-loop DWS measurements showed an approximately twofold reduction in RMS pointing jitter over 3--70 mHz in both directions. Furthermore, the nested control reduced TTL coupling by at least one order of magnitude below 5 mHz and by at least two orders below 0.5 mHz.

Although demonstrated here in an on-axis configuration, the nested scheme is also applicable to off-axis architectures such as the GRACE-FO LRI. Importantly, it retains the GRACE-FO-like DWS-FSM architecture while adding low-frequency SAA feedback, making it directly compatible with NGGM \cite{daras2024next,cesare2022drag}. By suppressing spacecraft attitude fluctuations, the scheme not only reduces TTL below other projected low-frequency noise sources but may also mitigate attitude-dependent accelerometer errors. This reduced noise floor allows future improvements in LFN and ACN to translate directly into improved mission sensitivity. The concept is also applicable to other space-based laser-interferometry missions requiring precision pointing \cite{el2020aedge,quanz2022large}.

Measurements were performed in air without thermal control, with the hexapod rotation limited to tens of microradians. These constraints left a residual TTL noise of $\sim2~\mathrm{nm}/\sqrt{\mathrm{Hz}}\times\mathrm{NSF}(f)$. Further improvement will require higher-precision rotational control and measurements under vacuum with improved thermal stability.

\begin{acknowledgments}
The authors acknowledge financial support by the Max Planck Society (MPG) (QOP18098), the Deutsche Forschungsgemeinschaft (DFG) (434617780-SFB 1464), the Relativistic Geodesy 1128, the Clusters of Excellence (EXC2123, EXC2122), and the German Aerospace Center (DLR) (50OQ2301).
\end{acknowledgments}	

\appendix
\section{End Matter}\label{end_matter}
\begin{figure}[!htp]
	\centering
	\includegraphics[width=0.48\textwidth]{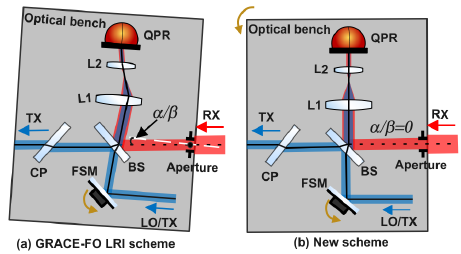}
	\caption{Comparison between (a) the GRACE-FO LRI pointing control scheme and (b) the nested pointing control scheme. In both configurations, the GRACE-FO optical bench is employed. $\alpha$ and $\beta$ represent the spacecraft attitude angles in yaw and pitch directions, respectively. The angles ($\alpha$/$\beta$) illustrated in (a) are exaggerated for visualization, whereas the angles are much smaller in a real scenario. L, lens; FSM, fast steering mirror; QPR, quadrant photoreceiver; BS, beamsplitter; CP, compensation plate; RX, receiving beam; TX, transmitting beam; LO, local beam.}
	\label{Scheme_comparison}
\end{figure}

\textit{Concept}---Figure~\ref{Scheme_comparison}(b) illustrates the proposed nested active pointing control for interspacecraft laser interferometry, while Fig.~\ref{Scheme_comparison}(a) shows the GRACE-FO LRI pointing scheme for comparison. In the GRACE-FO LRI, the spacecraft attitude is independently maintained by the AOCS, while DWS-FSM loops compensate for the residual pointing jitter. Because the optical bench remains fixed to the spacecraft, residual SAA variations ($\alpha$/$\beta$) change the RX-beam incidence angle and, consequently, the internal optical path. In the nested scheme, additional SAA-AOCS loops suppress these SAA variations by rotating the entire optical bench toward the RX beam, thereby maintaining near-normal incidence and keeping the FSM close to its central position. The faster DWS-FSM loops compensate residual beam misalignment during AOCS actuation to maintain RX-TX beam coalignment. Compared with the GRACE-FO LRI, where the FSM provides several milliradians of angular steering range \cite{abich2019orbit}, the nested scheme keeps the FSM near zero deflection, allowing a smaller steering range or greater margin with the same FSM.

The proposed SAA-AOCS loop requires no dedicated attitude-control thrusters, but a sufficiently precise SAA measurement and access to the existing spacecraft attitude actuators through an additional low-frequency feedback channel. Within the loop bandwidth, attitude control noise is suppressed by the loop gain, while the achievable performance is limited by the SAA measurement noise \cite{wei2026beam}. In our experiment, the SAA was inferred from the FSM-PSS measurements. A similar capability based on eddy-current sensors was implemented in the GRACE-FO LRI FSM \cite{abich2015grace}. The concept is particularly compatible with NGGM, whose 20~µrad attitude-control accuracy exceeds the 80~µrad rotational accuracy of the hexapod used in our experiment \cite{daras2024next,cesare2022drag,massotti2021next}. For missions with more stringent attitude pointing requirements, the SAA could instead be measured with a higher-precision interferometric angular sensor based on DWS.

\begin{figure}[!htp]
	\centering
	\includegraphics[width=0.483\textwidth]{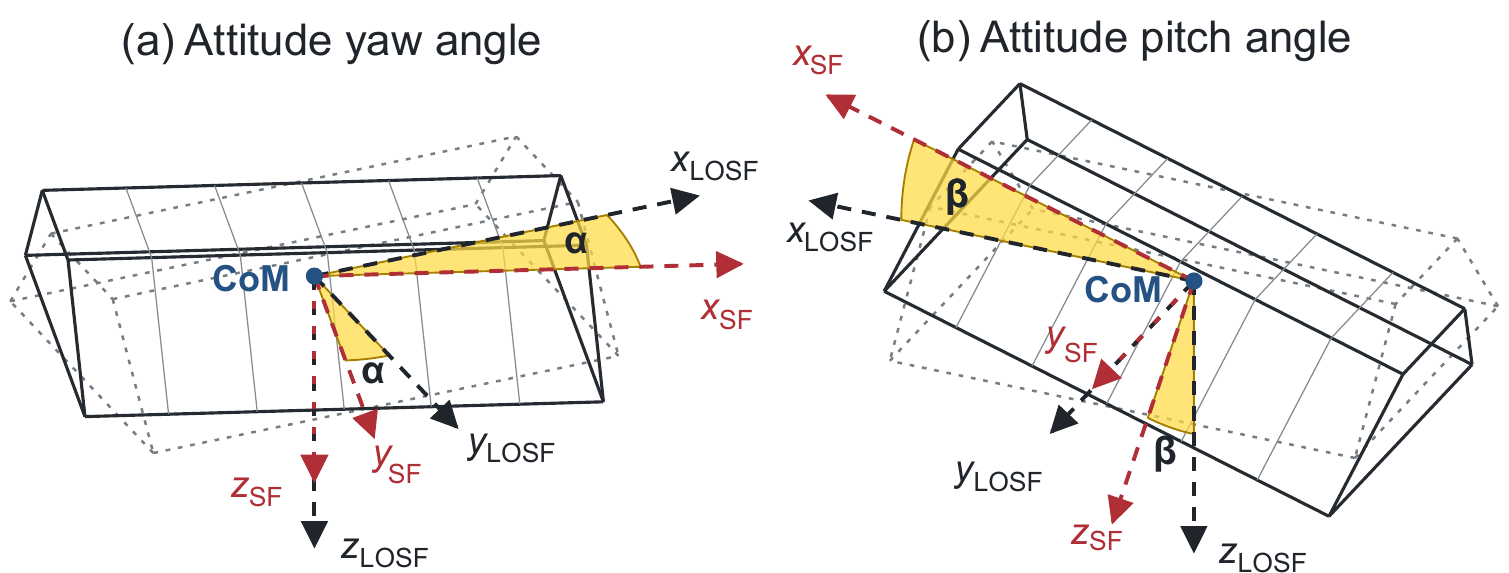}
	\caption{Schematic illustrating the definitions of the spacecraft attitude (a) yaw and (b) pitch angles. The spacecraft frame (SF) and line-of-sight frame (LOSF) are indicated. CoM denotes the center of mass.}
	\label{angle_definition}
\end{figure}

\textit{Angle definition}---The SAA is defined in Figs.~\ref{angle_definition} with respect to the LOS, which, in GRACE-like missions, is defined by the line connecting the centers of mass of the two spacecraft. The yaw angle $\alpha$ and pitch angle $\beta$ denote rotations about the $z$ and $y$ axes of the LOS frame, respectively. On the optical bench, the corresponding incidence angles of the RX beam at the aperture are equivalently described by $\alpha$ and $\beta$. The FSM angular deflections in yaw and pitch are denoted by $\gamma$ and $\eta$, respectively, and are obtained from the PSS measurements. Within the DWS--FSM loops bandwidth, the attitude angles $\alpha$ and $\beta$ can be derived from the measured FSM deflections $\gamma$ and $\eta$ as 
\begin{equation}
\begin{pmatrix} \alpha \\ \beta \end{pmatrix}
= M\begin{pmatrix} \gamma \\ \eta \end{pmatrix},
\label{fsm_to_attitude}
\end{equation}
where $M$ is the transfer matrix from the FSM frame to the spacecraft frame, incorporating the effects of the imaging optics and beam reflection. In the experiment, $M$ is calibrated using controlled FSM rotations and the corresponding angular measurements of the TX beam.

\begin{figure}[!htp]
	\centering
	\includegraphics[width=0.48\textwidth]{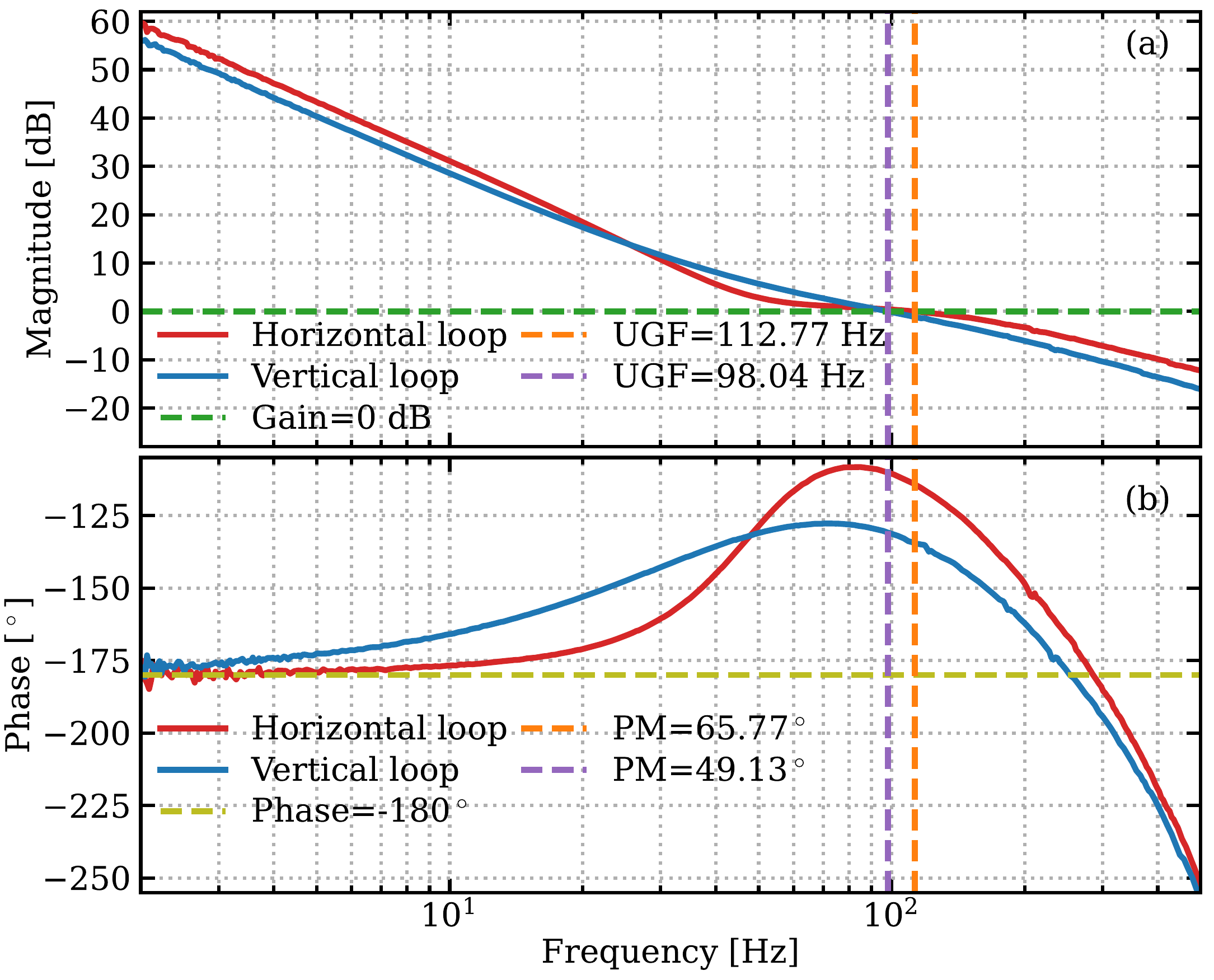}
	\caption{Measured open-loop transfer functions of the DWS-FSM loops: (a) magnitude and (b) phase responses. The horizontal and vertical loops are driven by the corresponding DWS signals. The discrepancy between the two loops arises from differences in the coupling factors from the FSM angle to the DWS signal, as well as differences in the servo system gain factors. UGF, unity gain frequency; PM, phase margin.}
	\label{tf_DWS_fsm_loop}
\end{figure}

\begin{figure}[!htp]
	\centering
	\includegraphics[width=0.483\textwidth]{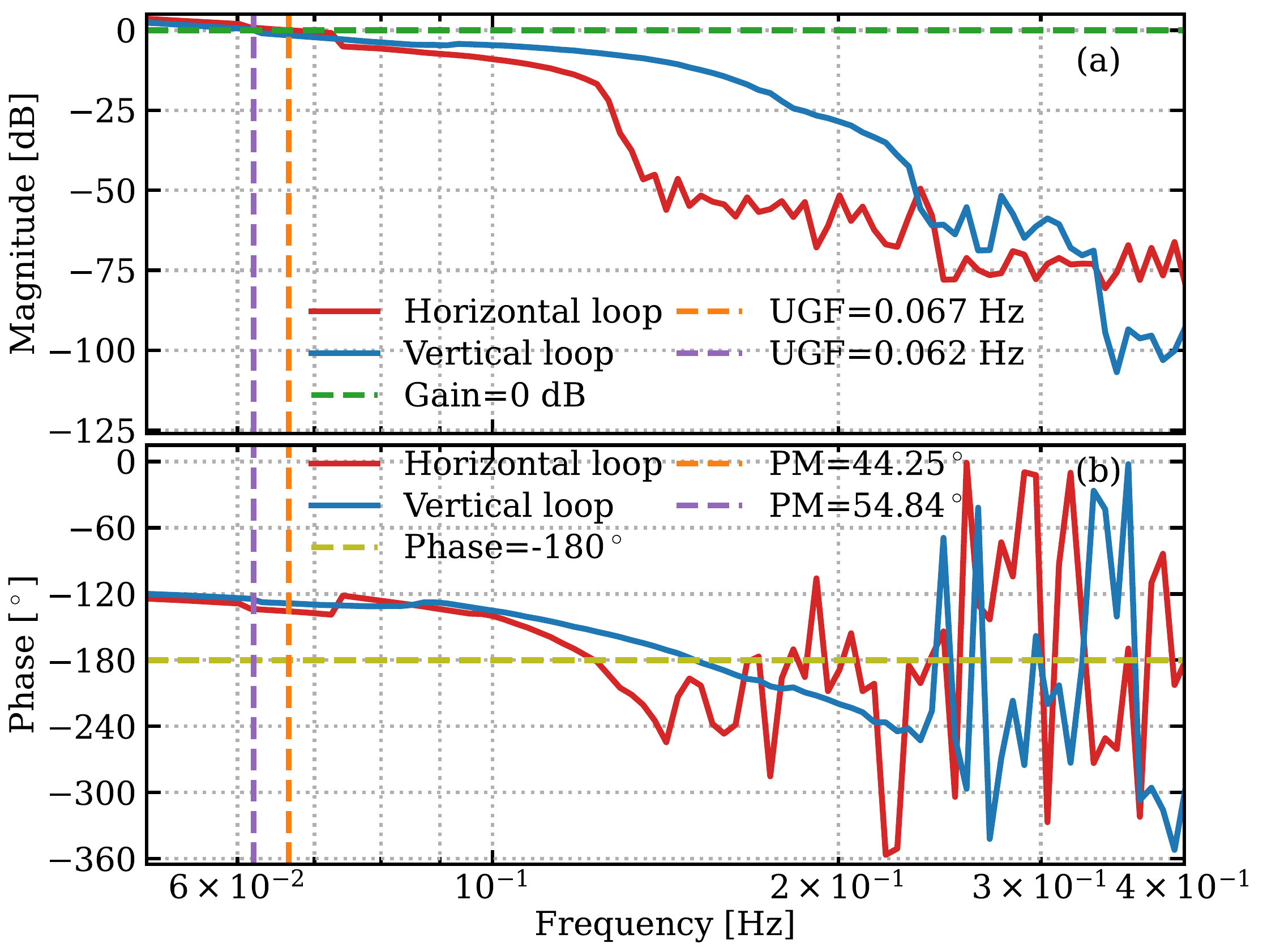}
	\caption{Measured open-loop transfer functions of the SAA-AOCS loops: (a) magnitude and (b) phase responses. The horizontal and vertical loops represent the feedback loops driven by the $\gamma$ (yaw) and $\eta$ (pitch) FSM angular deflections, respectively. UGF, unity gain frequency; PM, phase margin.}
	\label{tf_saa_aocs_loop}
\end{figure}

\textit{Transfer functions}---We measured the open-loop transfer functions of the DWS-FSM and SAA-AOCS loops with a dynamic signal analyzer (Stanford Research Systems SR785), injecting swept-sine perturbations at the FSM setpoint for the DWS-FSM loop and at the $\gamma$ or $\eta$ signal, as defined in Fig.~\ref{Pointing_loops}, for the SAA-AOCS loop. As shown in Fig.~\ref{tf_DWS_fsm_loop}, the horizontal and vertical DWS-FSM loops  achieve unity-gain bandwidths of 112.77~Hz and 98.04~Hz, with phase margins of 65.77$^{\circ}$ and 49.13$^{\circ}$, respectively. The SAA-AOCS loops, shown in Fig.~\ref{tf_saa_aocs_loop}, exhibit unity-gain bandwidths of 67~mHz (horizontal) and 62~mHz (vertical), with phase margins of 44.25$^{\circ}$ and 54.84$^{\circ}$, respectively.

\begin{table*}
\caption{\label{tab:NSF}
Noise levels and corresponding noise-shaping functions. LFN and ACN denote laser-frequency noise and accelerometer noise, respectively.}
\begin{ruledtabular}
\begin{tabular}{lll}
Quantity & Number & $\mathrm{NSF}(f)$ \\
\hline
GRACE-FO LRI requirement \cite{abich2019orbit} & $80~\mathrm{nm}/\sqrt{\mathrm{Hz}}$ &
$\sqrt{1+\left(3~\mathrm{mHz}/{f}\right)^2}
\sqrt{1+\left(10~\mathrm{mHz}/{f}\right)^2}$
\\[1.2ex]
Reference level and NGGM goal \cite{nicklaus2022towards} & $2~\mathrm{nm}/\sqrt{\mathrm{Hz}}$,$20~\mathrm{nm}/\sqrt{\mathrm{Hz}}$ &
$\sqrt{\left(1+1~\mathrm{mHz}/{f}\right)^2
      \left(1+10~\mathrm{mHz}/{f}\right)^2}$
\\[1.2ex]
Pointing  & $100~\mathrm{nrad}/\sqrt{\mathrm{Hz}}$, 10 µrad$/\sqrt{\mathrm{Hz}}$ &
$\sqrt{\left(1+1~\mathrm{mHz}/{f}\right)^2
      \left(1+10~\mathrm{mHz}/{f}\right)^2}$
\\[1.2ex]
LFN-NGGM \cite{nicklaus2020laser} & $40~\mathrm{Hz}/\sqrt{\mathrm{Hz}}$ &
$\sqrt{1+\left(10~\mathrm{mHz}/{f}\right)^2}$
\\[1.2ex]
ACN--GRACE-FO \cite{darbeheshti2017instrument} &$10^{-10}~\mathrm{m/s^{2}/\sqrt{Hz}}$ &
$\sqrt{1+5~\mathrm{mHz}/{f}}$
\\[1.2ex]
ACN--NGGM \cite{QSG4EMTConsortium2024} & $5 \times 10^{-12}~\mathrm{m/s^{2}/\sqrt{Hz}}$&
$\sqrt{1+\left(1~\mathrm{mHz}/{f}\right)^2+\left(f/{10^{-1} \mathrm{Hz}}\right)^4}$
\\
\end{tabular}
\end{ruledtabular}
\end{table*}

\textit{Noise-shaping functions and equivalent noise}---Table~\ref{tab:NSF} lists the NSFs adopted for the ranging, pointing, LFN, and ACN ASDs. For the TTL-induced ranging noise in Fig.~\ref{estimated_TTL_diff_acc}, we assume identical attitude fluctuations on the two spacecraft and use the flight-derived GRACE-FO LRI yaw and pitch coupling factors $a_1=45.29$ µm/rad, $b_1=94.52$ µm/rad for spacecraft 1, and $a_2=136.47$ µm/rad, $b_2=86.94$ µm/rad for spacecraft 2~\cite{wegener2020tilt,wegener2022analysis}. The equivalent LFN contribution is estimated using the wavelength of 1064 nm and the NGGM interspacecraft separation of 100 km \cite{nicklaus2020laser}. The ACN values in Table~\ref{tab:NSF} are single-spacecraft values. Assuming independent accelerometer noise on the two spacecraft, the corresponding differential-acceleration ASD includes a factor of $\sqrt{2}$.

\nocite{*}

\bibliography{sample}

\end{document}